\documentclass[journal,comsoc]{IEEEtran}
\usepackage{cite}
\usepackage{graphicx}
\usepackage{psfrag}  
\usepackage[table]{xcolor}
\usepackage{subfig}

\usepackage{mathtools}
\usepackage{amsfonts}
\usepackage{amssymb}
\usepackage[cmintegrals]{newtxmath}
\DeclareMathAlphabet\mathbfcal{OMS}{cmsy}{b}{n}
\usepackage{bm}

\usepackage{acronym}
\newacro{PLC}{power line communication}
\newacro{RF}{radio frequency}
\newacro{VLC}{visible light communication}
\newacro{SNR}{signal-to-noise ratio}
\newacro{nSNR}{normalized SNR}
\newacro{PSD}{power spectral density}
\newacro{CFR}{channel frequency response}
\newacro{DFT}{discrete-time Fourier transform}
\newacro{pdf}{probability density function}
\newacro{CIR}{channel impulse response}
\newacro{CSI}{channel state information}
\newacro{PLS}{physical layer security}
\newacro{OA}{optimal power allocation}
\newacro{UA}{uniform power allocation}
\newacro{AWGN}{additive white Gaussian noise}
\newacro{BER}{bit error rate}
\newacro{BPSK}{binary phase shift keying}
\newacro{LGRC}{linear Gaussian relay channel}
\newacro{LGC}{linear Gaussian channel}
\newacro{CGRC}{circular Gaussian relay channel}
\newacro{MIMO}{multiple-input multiple-output}

\begin{document}
	\title{In-Home Broadband PLC Systems Under the Presence of a Malicious Wireless Device: Physical Layer Security Analysis}
	
	\author{\^Andrei Camponogara,~H. Vincent Poor, \textit{Fellow,~IEEE}, and
		Mois\'es Vidal Ribeiro, \textit{Senior~Member,~IEEE}
		% <-this % stops a space
		\thanks{The authors would like to thank CAPES (Finance Code 001), CNPq, INERGE, FAPEMIG for their financial support. This study was supported in part by the U.S. National Science Foundation under Grants CCF-0939370 and CCF-1513915. Finally, the authors also would like to thank professor Richard Demo Souza for his contributions in this study.}
		\thanks{\^Andrei Camponogara is with the Department of Electrical Engineering, Federal University of Juiz de Fora, Juiz de Fora, MG 36036 330, Brazil (e-mail: andrei.camponogara@engenharia.ufjf.br).}
		\thanks{H. V. Poor is with the School of Engineering and Applied Science, Princeton University, Princeton, NJ 08544 USA (e-mail: poor@princeton.edu).}
		\thanks {Mois\'es V. Ribeiro is with the Department of Electrical Engineering, Federal University of Juiz de Fora and also with the Smarti9 Ltd., Juiz de Fora, MG 36036 330, Brazil (e-mail: mribeiro@engenharia.ufjf.br).}}
	
\markboth{}%
{Shell \MakeLowercase{\textit{et al.}}: Bare Demo of IEEEtran.cls for IEEE Communications Society Journals}

\maketitle

\begin{abstract}
	This paper investigates the physical layer security (PLS) of a broadband in-home power line communication (PLC) system when a malicious wireless device tries to eavesdrop private messages exchanged between two PLC devices. Such a security issue arises when electric power circuits, which are used for data communication, are constituted by unshielded power cables. 
	In this regard, the hybrid wiretap channel model for formulating achievable secrecy rate and secrecy outage probability is considered. Additionally, a data set of channel estimates and measured additive noises obtained from a measurement campaign carried out in several in-home facilities is used for providing practical results, which can offer direction for dealing with the security aspects of broadband in-home PLC systems in the physical layer perspective. The attained results show high values of secrecy outage probability for all chosen values of target secrecy rate and total power transmission (practical and theoretical) when the PLC devices are far from each other and the eavesdropper is close to the PLC transmitter (i.e., the distance is shorter than 2 meters). Overall, the numerical results show that the vulnerability of broadband in-home PLC systems, in terms of PLS, is relevant when practical values of total transmission power apply. Therefore, a rethinking  of the use of unshielded power cables or new designs of the broadband in-home PLC system deserves special attention for ensuring security at the physical layer.		
\end{abstract}


\IEEEpeerreviewmaketitle

% --------------------------------------------------------------
% INTRODUCTION
% --------------------------------------------------------------
\section{Introduction}
\IEEEPARstart{R}{ecently}, \ac{PLC} systems have been widely studied by both academic and business sectors since electric power infrastructures are pervasive and the connections to them are ubiquitous. Moreover, they support data communication for indoor (residential and commercial buildings) and outdoor (low- and medium-voltages) electric power grids \cite{moises, facina, luis}. Currently, transportation systems (e.g., car, ship, train, spacecraft and aircraft) \cite{mohammadi,ships,trains,spacecraft,andrei} have constituted a new frontier for \ac{PLC} systems. Despite the well-known advantages for data communication, such as omnipresence, low-cost implementation, and easy installation when the voltage level is low (i.e., below $400$~V), electric power systems (generation, transmission, and distribution) were initially designed and deployed for energy delivery rather than data communication. Therefore, data carrying signals traveling through electric power grids may suffer severe degradation and attenuation due to the long distance between source and destination nodes; the use of high frequencies for data communication in a non-ideal conductor (i.e., power cables); the presence of multipath effects due to the existence of impedance mismatching at the branching points and load's connection points; the noise generated by the dynamic of loads (e.g., consumers and utilities) \cite{cortes2005,cortes2011,cataliotti,huang,facina,oliveira2017}; coupling losses in the injecting and extracting circuits belonging to the front/end of \ac{PLC} modems \cite{janse,luis}, and interference with other telecommunication systems operating in the same frequency band because power cables are usually unshielded\footnote{Aerial unshielded power cables are used in almost all electric power systems. Underground and shielded are used in high-density downtown areas.}.

In \cite{oliveira2015,oliveira2016} were shown that the unshielded characteristic of power cables is an opportunity for having a hybrid \ac{PLC}-wireless system\footnote{PLC-wireless means PLC and wireless channels are used in cascade or concatenated.}, in which \ac{PLC} and wireless devices operating in the same frequency band can communicate with each other. On the other hand, the broadcast nature of electric power grids and the fact that these grids are mainly constituted by unshielded power cables can potentially turn \ac{PLC} systems vulnerable to malicious users \cite{oliveira2015,oliveira2016}. As a matter of fact, \ac{PLC} signals (i.e., data carrying signals) flowing through an unshielded power cable radiates electromagnetic fields that can be sensed by a wireless device operating in the same frequency band and located in the vicinity of the electric power circuit. If the wireless device is a malicious one, then the investigation of \ac{PLS} for \ac{PLC} systems becomes a very appealing one to quantify the lack of security in these media and, as a consequence, to circumvent the leaking information from the \ac{PLC} system to a wireless device.

The idea of analyzing security at the physical layer level was brought to the attention of the academic community in 1970's  \cite{wyner,cheong,csiszar}. In particular, \cite{wyner} introduced the degraded wiretap channel. In the sequel, \cite{cheong} evaluated the secrecy capacity for the Gaussian wiretap channel and \cite{csiszar} proposed a more general model known as the non-degraded wiretap channel. Recently, fading wiretap channels \cite{parada,liang,liang2} and \ac{MIMO} wiretap channels \cite{oggier,shafiee,mukherjee,shlezinger,yang}, among others, have been investigated. For a review of this area, see \cite{poor}.

According to the literature, few investigations have assessed the security of \ac{PLC} systems under the \ac{PLS} perspective \cite{salem,shafie,andrei2}. \cite{salem} analyzed the secrecy capacity for both \ac{PLC} and hybrid PLC/wireless\footnote{PLC/wireless means PLC and wireless channels are used in parallel.} single relay channels in the presence of a passive eavesdropper. Whereas \cite{shafie} introduced an artificial noise scheme to improve the security of the hybrid PLC/wireless channel. Last, \cite{andrei2} assessed \ac{PLS} of the low-bit-rate hybrid PLC/wireless channel and its incomplete versions by adopting achievable ergodic secrecy rate and secrecy outage probability. Nonetheless, these studies have not considered \ac{PLC} signals radiated from the power cable to the air and accessed by a malicious wireless device and discussed analyses based on the use of real data set to quantify \ac{PLS} in broadband in-home \ac{PLC} systems\footnote{The term broadband PLC system was introduced by the PLC community and owns a different meaning in comparison to the telecommunication community. It refers to high-data-rate PLC systems that cover the frequency bands $1.7-30$ MHz, $1.7-50$ MHz, and $1.7-86$ MHz.}. According to the authors understanding, this problem may be entirely portrayed by the well-known in-home hybrid PLC-wireless channels \cite{oliveira2015,oliveira2016} since they correctly characterize this environment.
These studies discussed measurement campaign, statistical characterization, modeling, and analysis of measured channels associated with the existing link between \ac{PLC} and wireless devices (i.e., hybrid PLC-wireless channel), which are supposed to be operating in the same frequency band ($1.7-100$ MHz). Additionally, \cite{oliveira2015} presented the in-home \ac{PLC} channels obtained in the same measurement campaign and compared them with the respective hybrid \ac{PLC}-wireless counterparts, which constitutes relevant information for differentiating these channels. 

\begin{figure*}[!htb]
	\centering
	\psfrag{pl}[ll][ll][.65]{\textit{Power Line}}
	\includegraphics[width=16cm]{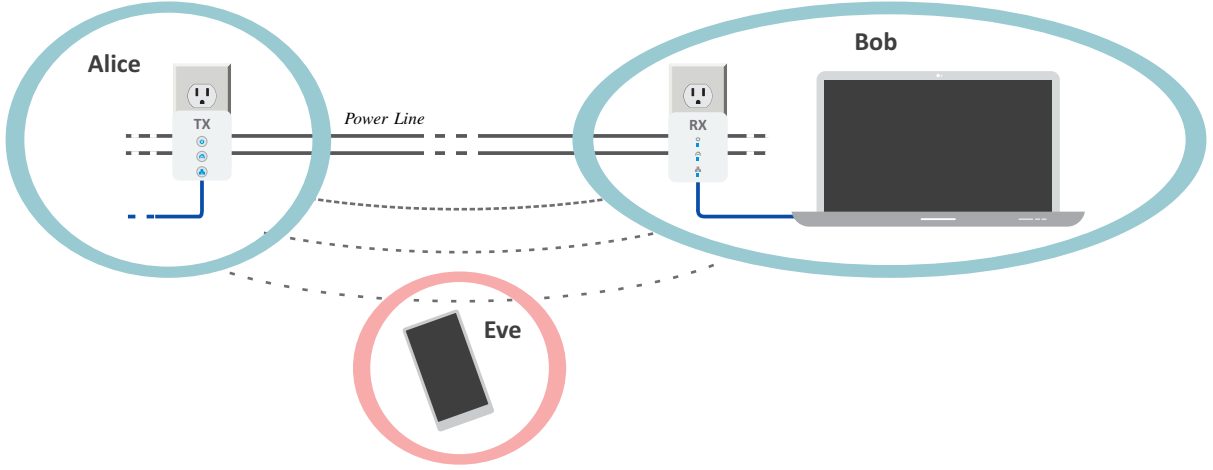}
	\caption{Illustration of broadband data communication between two \ac{PLC} devices under the presence of a malicious wireless device.}
	\label{fig:scenario}
\end{figure*}

It is essential to point out that such in-home hybrid \ac{PLC}-wireless channels may represent the scenario where a malicious wireless device threatens the security of a broadband PLC system, see Fig. \ref{fig:scenario}. The \ac{PLC} transmitter (Alice) sends private messages to the legitimate \ac{PLC} receiver (Bob) whereas a malicious wireless device (Eve) eavesdrops the information carried by the \ac{PLC} signal from its radiated component, which is a consequence of the propagation of the \ac{PLC} signal through an unshielded cable. This scenario is termed hybrid wiretap channel. Observe that, in this kind of channel, Alice can not realize that Eve is overhearing private messages sent to Bob, which is a typical situation associated with the use of unshielded power cables.

In this regard, this study aims to quantitatively discuss important issues related to \ac{PLS} when the hybrid wiretap channel model, the frequency band $1.7-86$~MHz (in agreement with ITU-T Rec. G.9964), and the use of a data set constituted by estimates of the \ac{CFR} and measured additive noises are taken into account. The main contributions of this study are as follows:
\begin{itemize}
	\item Introductions of the hybrid wiretap channel model and mathematical formulations of achievable secrecy rate and secrecy outage probability. Also, discussion of numerical results related to secrecy outage probability and ergodic achievable secrecy rate considering a passive eavesdropper (i.e., the \ac{CSI} of Eve is not available at Alice). Numerical results make use of a real data set obtained from the measurement campaign carried out in Brazilian in-home facilities \cite{oliveira2015,oliveira2016}.
	\item Performance analysis considering the following issues: distinct distances between Alice and Bob; different positions of Eve in relation to Alice and Bob; distinct levels of the total transmission power (e.g., practical and theoretical); and two types of resource allocation technique (optimal and uniform). Such an analysis allows to quantify how much a wireless device can threaten the security of a broadband \ac{PLC} system in practice.
\end{itemize}

Based on the numerical analysis, it is clear that the vulnerability of the broadband in-home \ac{PLC} system is relevant for practical values of total transmission power when unshielded power cables are part of the electric power grid and a malicious wireless device is near the \ac{PLC} transmitter. In the worst scenario, in which Bob is far from Alice and Eve is close to Alice (i.e., the distance is shorter than $2$~meters), the secrecy outage probability is high regardless of the total transmission power (practical and theoretical values) or the adopted target secrecy rate. In other words, the level of security at the physical layer of the broadband in-home \ac{PLC} system may be close to zero. Therefore, a rethinking of the use of unshielded power cables or the introduction of novelties in broadband in-home PLC systems has to be carried out for ensuring security at the physical layer.
	
The rest of this paper is organized as follows: Section \ref{Formulation} presents the hybrid wiretap channel model; Section \ref{Capacity} describes mathematical formulations for achievable secrecy rate and secrecy outage probability; Section \ref{Results} shows the numerical results; and, finally, Section \ref{Conclusion} states some concluding remarks.

\subsubsection*{Notation}
Lower-case and upper-case boldface symbols denote vectors in the discrete-time and -frequency domains, respectively. $\mathbf{\Lambda}_D \triangleq \text{\textbf{diag}}\{D[0],~D[1],\cdots,~D[N-1]\}$ represents the diagonal matrix of the $N$-length random vector $\mathbf{D}$. $\mathbf{R}_{DD} = \mathbb{E}[\mathbf{D}\mathbf{D}^\dagger]$ is the auto-correlation matrix of the random vector $\mathbf{D}$. $\mathbfcal{F}$ is the $N$-size and normalized \ac{DFT} matrix. $\mathbf{0}_{N\times1}$ is the $N$-length column vector of zeros. $\mathbf{I}_N$ is the $N\times N$ identity matrix. $\det(\mathbf{\Lambda}_D)$ denotes the determinant of the matrix $\mathbf{\Lambda}_D$. $h(\cdot)$ refers to the entropy function. $|\cdot|$ denotes the modulus operator. $\|\cdot\|$ is the Euclidean norm. $\mathbb{E}[\cdot]$ is the expectation operator. $\{\cdot\}^T$ denotes the transpose operator. $\{\cdot\}^\dagger$ is the Hermitian operator. $\textbf{\text{tr}}(\cdot)$ is the trace operator. $\mathbb{P}\{c<d\}$ is the probability that $c$ is less than $d$.

% --------------------------------------------------------------
% PROBLEM FORMULATION
% --------------------------------------------------------------
\section{Problem Formulation}\label{Formulation}
\ac{PLS} is an interesting issue for investigation in \ac{PLC} systems since in the majority of the cases, the power cables are unshielded and densely interconnected in low-voltage levels. Also, they are easy to connect and access physically. From the author point of view, a fundamental question about \ac{PLS} in broadband \ac{PLC} systems context is how secure these systems can be related to the presence of malicious device connected to the power line (e.g., the eavesdropper is another \ac{PLC} device operating in the same frequency band) or wirelessly connected to the power line (e.g., the eavesdropper is a wireless device operating in the same frequency band and the vicinity of the power line). Regarding the latter form of eavesdropping, this section formulates the problem to investigate \ac{PLS} of a broadband \ac{PLC} system, which operates in in-home facilities, when a malicious wireless device is located in the vicinity to overhear the private messages exchanged between two \ac{PLC} devices.

\begin{figure}[!htb]
	\centering
	\psfrag{hb}[c][c][.8]{$h_{B}[n,m]$}
	\psfrag{he}[c][c][.8]{$h_{E}[n,m]$}
	\psfrag{x}[c][c][.8]{$x[n]$}
	\psfrag{y}[c][c][.8]{$y_{B}[n]$}
	\psfrag{z}[c][c][.8]{$y_{E}[n]$}
	\psfrag{vb}[l][l][.8]{$v_{B}[n]$}
	\psfrag{ve}[l][l][.8]{$v_{E}[n]$}
	\includegraphics[width=8.5cm]{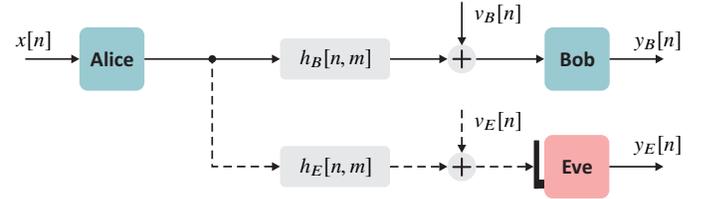}
	\caption{Block diagram of the hybrid wiretap channel model. The continuous and dashed lines represent the power line the hybrid PLC-wireless links, respectively.}
	\label{fig:Hybrid}
\end{figure}

In this regard, the block diagram in Fig. \ref{fig:Hybrid} shows the investigated hybrid wiretap channel model. A \ac{PLC} transmitter Alice ($A$) sends private messages to the legitimate \ac{PLC} receiver Bob ($B$). Meanwhile, a malicious wireless device Eve ($E$) located in the vicinity of Alice or Bob overhears part of the private messages. This scenario may happen due to an inherent characteristic of unshielded power cables: they radiate part of the \ac{PLC} signal flowing through them. In this scenario, Eve is operating in the same frequency band of Alice and Bob and she is capable of sensing the radiation of the \ac{PLC} signal without Alice realizes the presence of Eve. In accord with \cite{oliveira2015,oliveira2016}, the wireless propagation of the radiated \ac{PLC} signal can be covered by the broadband hybrid \ac{PLC}-wireless channels. Also, the respective broadband in-home \ac{PLC} channels presented in \cite{oliveira2015} can represent the link between Alice and Bob.

Considering that \ac{PLC} and hybrid \ac{PLC}-wireless channels are time-varying systems, one can state that $\{h_l[n,m]\}$, where $l\in\{B,~E\}$, denotes the discrete-time version of the time-varying channels associated with Alice-Bob and Alice-Eve links, respectively. Based on this assumption, the discrete-time representation of the received signal at the input of the $l^{th}$ receiver is given by
\begin{align}
	y_l[n] = \sum_{m=-\infty}^{\infty} Ax[m]h_l[n,m] + v_l[n],
\end{align}
where $\{x[n]\}$ is constituted by an infinite number of $N$-length symbols ($N$-block symbols) and refers to the transmitted sequence; $A \in \mathbb{R}_+$ is the amplitude of the transmitted sequence; $h_l[n,m]$ denotes \ac{CIR} seen by the $l^{th}$ receiver in the $n^{th}$ sample when an impulse is injected in the $m^{th}$ sample by Alice; and $\{v_l[n]\}$ denotes the additive noise sequence. It is assumed that both $\{x[n]\}$ and $\{v_l[n]\}$ are independent and wide-sense stationary random processes. 
	
Assuming that the time interval associated with an $N$-block symbol is shorter than the coherence time of the \ac{PLC} and hybrid \ac{PLC}-wireless channels in the continuous time-domain, then these channels can be considered linear and time-invariant during a time interval corresponding to an $N$-block symbol. In this regard, the discrete-time \ac{CIR} during a time interval corresponding to an $N$-block symbol is time-invariant and, as a consequence, $h_l[m,n]=h_l[n-m]$, such that $\{\text{h}_l[n]\}_{n=0}^{L_l-1}$, where $L_l$ is the length of \ac{CIR} associated with the link between Alice and the $l^{th}$ receiver. The vectorial representation of the discrete-time version of such channels during one $N$-block symbol duration is $\mathbf{h}_l~=~[\text{h}_l[0],~\text{h}_l[1],~\ldots,~\text{h}_l[L_l-1]]^T$, while $\mathbf{H}_l~=~[{H}_l[0],~{H}_l[1],~\ldots,~{H}_l[N-1]]^T$ denotes its vectorial representation in the frequency domain, where $\mathbf{H}_l = \mathbfcal{F}[\mathbf{h}_l^T,~\mathbf{0}_{N-L_l}^T]^T$ and $N$ denotes the number of sub-channels. Hereafter, the diagonal matrices $\mathbf{\Lambda}_{\mathbfcal{H}_l}~=~\text{\textbf{diag}}\left\lbrace {H}_l[0],~{H}_l[1],~\ldots,~{H}_l[N-1]\right\rbrace$ and
$\mathbf{\Lambda}_{|\mathbfcal{H}_l|^2} = \text{\textbf{diag}}\{|{H}_l[0]|^2,~|{H}_l[1]|^2,~\cdots,|{H}_l[N-1]|^2\}$ will be used. 

Moreover, the vectorial representation of the $N$-block 
symbol, which is defined in the frequency domain for transmission, is $\mathbf{X}\in\mathbb{C}^{N\times1}$ so that $\mathbb{E}[\mathbf{X}] = \mathbf{0}_{N\times1}$ and $\mathbb{E}[\mathbf{X}\mathbf{X}^\dagger] = N\mathbf{\Lambda}_{P}$, where $\mathbf{\Lambda}_{P} = \text{\textbf{diag}}\left\lbrace P[0],~P[1],~\ldots,~P[N-1]\right\rbrace$ is the matrix representation of the allocated power and $\textbf{\textbf{\text{tr}}}(\mathbf{\Lambda}_{P}) = P_T$ is the total transmission power. Also,  $\mathbf{V}_{l}\in\mathbb{C}^{N\times1}$ is the frequency domain vectorial representation of the zero mean additive noise, such that $\mathbb{E}[\mathbf{V}_l] = \mathbf{0}_{N\times1}$ and $\mathbb{E}[\mathbf{V}_l^{\phantom \dagger}\mathbf{V}_l^\dagger] = N\mathbf{\Lambda}_{P_{V_l}}$, in which $\mathbf{\Lambda}_{P_{V_l}} = \text{\textbf{diag}}\left\lbrace P_{V_l}[0],~P_{V_l}[1],~\ldots,P_{V_l}[N-1]\right\rbrace$ and $P_{V_l}[k]$ is the additive noise power in the $k^{th}$ sub-channel. 

Given the aforementioned formulation, the following twofold questions arise:
\textit{How secure is a broadband \ac{PLC} system at the physical layer level when Eve is a wireless device operating in the same frequency band and she is in the vicinity of Alice or Bob in a practical perspective? In other words, how much information is leaked to Eve and how it impacts the secrecy outage probability?}
Aiming to answer these questions, Section \ref{Capacity} deduces the achievable secrecy rate and secrecy outage probability for the hybrid wiretap channel model and Section \ref{Results} discusses numerical results that support important findings related to \ac{PLS} in \ac{PLC} systems.
	
% ---------------------------------------------------------
% SECRECY OUTAGE PROBABILITY
% ---------------------------------------------------------
\section{Achievable Secrecy Rate and Secrecy Outage Probability}\label{Capacity}
This section deduces mathematical expressions to compute the achievable secrecy rate and secrecy outage probability for the hybrid wiretap channel model. To do so, similar to the \ac{LGRC} addressed in \cite{mitra}, \ac{PLC} and hybrid \ac{PLC}-wireless channels are assumed $N$-block linear Gaussian channels with finite memory (i.e., $L_{\max} = \max\limits_{l}L_{l}$). However, the inter-block interference caused by the memory of \acp{CIR} and the correlated noises make difficult to evaluate the achievable data rate \cite{goldsmith}. On the other hand, the proposal in \cite{goldsmith} can overcome this drawback. It states that the $N$-block \ac{CGRC} eliminates the inter-block interference when $N\gg L_{\max}$. Besides, \ac{LGRC} tends to $N$-\ac{CGRC} as $N\rightarrow\infty$. Hence, $N$-\ac{CGRC} channels model \ac{PLC} and hybrid \ac{PLC}-wireless channels since $N\rightarrow\infty$.
	
Assuming perfect synchronization, the vectorial representation, in the frequency domain, of the received symbol at the $l^{th}$ receiver is given by
\begin{align}
	\mathbf{Y}_l = \mathbf{\Lambda}_{\mathbfcal{H}_l}\mathbf{X} + \mathbf{V}_l.
\end{align}
Then the mutual information between Alice and the $l^{th}$ receiver can be expressed as
\begin{align}\label{eq:Ixy}
	I(\mathbf{X};\mathbf{Y}_{l}) &= h(\mathbf{Y}_{l}) - h(\mathbf{Y}_l|\mathbf{X}) \nonumber \\
								 &= h(\mathbf{Y}_l) - (h(\mathbf{\Lambda}_{\mathbfcal{H}_l}\mathbf{X}|\mathbf{X}) + h(\mathbf{V}_l|\mathbf{X}))  \nonumber \\
								 &= h(\mathbf{Y}_l) - h(\mathbf{V}_l).
\end{align}
	
Considering the additive noise and transmitted symbols as Gaussian random processes, the entropy of $\mathbf{Y}_l$ and $\mathbf{V}_l$ are given by
\begin{align}
	h(\mathbf{Y}_l) &= \log_2\left[(\pi e)^{N}\det(\mathbf{R}_{\mathbf{Y}\mathbf{Y},l})\right] \nonumber \\
\end{align}
and
\begin{align}
	h(\mathbf{V}_l) &= \log_2\left[(\pi e)^{N}\det(\mathbf{R}_{\mathbf{V}\mathbf{V},l})\right], \nonumber \\
\end{align}
respectively, in which $\mathbf{R}_{\mathbf{Y}\mathbf{Y},l} = \mathbf{\Lambda}_{\mathbfcal{H}_l}\mathbf{R}_{\mathbf{X}\mathbf{X}}\mathbf{\Lambda}_{\mathbfcal{H}_l}^\dagger + \mathbf{R}_{\mathbf{V}\mathbf{V},l}$. Thus, the capacity between Alice and the $l^{th}$ receiver can be expressed as
\begin{align}\label{eq:Ixy2}
	C_l &= \max_{f_\mathbf{X}(\mathbf{x}):\mathbf{tr}(\mathbf{R_{XX}})\leq NP_T} 
	     I(\mathbf{X};\mathbf{Y}_l) \nonumber \\
	    &= \max_{\mathbf{tr}(\mathbf{\Lambda}_{P})\leq P_T}\log_2\left[\det\left(\mathbf{I}_{N} + \mathbf{\Lambda}_{\gamma_l}\right)\right],
\end{align}
where $f_\mathbf{X}(\mathbf{x})$ is the joint density function of $\mathbf{X}$ and
\begin{align}
	\mathbf{\Lambda}_{\gamma_{l}}  &= \frac{\mathbf{\Lambda}_{\mathbfcal{H}_l}\mathbf{R}_{\mathbf{X}\mathbf{X}}\mathbf{\Lambda}_{\mathbfcal{H}_l}^\dagger}{\mathbf{R}_{\mathbf{V}\mathbf{V},l}} \nonumber \\ 
								   &= \mathbf{\Lambda}_{P}\mathbf{\Lambda}_{|\mathbfcal{H}_{l}|^2}\mathbf{\Lambda}_{P_{V_l}}^{-1}.
\end{align}
Hence, the secrecy capacity is given by \cite{jorswieck}
\begin{align}\label{eq:Cs}
	C_S = \max_{f_\mathbf{X}(\mathbf{x}):\mathbf{tr}(\mathbf{R_{XX}})\leq NP_T}\left[I(\mathbf{X};\mathbf{Y}_B) - I(\mathbf{X};\mathbf{Y}_E)\right]^+~[\text{bps/Hz}],
\end{align}
in which $\max[b]^+ = \max(0,b)$. In contrast, (\ref{eq:Cs}) is difficult to calculate. In accord with \cite{jorswieck}, a lower bound can be used as follows
\begin{align}\label{eq:Cs2}
	C_S &\geq\left[\max_{f_\mathbf{X}(\mathbf{x}):\mathbf{tr}\left(\mathbf{R}_{\mathbf{X}\mathbf{X}}\right)\leq NP_T}I(\mathbf{X};\mathbf{Y}_B) - \max_{f_\mathbf{X}(\mathbf{x}):\mathbf{tr}\left(\mathbf{R}_{\mathbf{X}\mathbf{X}}\right)\leq NP_T}I(\mathbf{X};\mathbf{Y}_E)\right]^+ \nonumber \\
        &=   \left[C_B - C_E\right]^+.
\end{align}

As discussed in \cite{yan}, a wiretap code is necessary to obtain the secrecy capacity. Further, for having perfect secrecy (i.e., the private messages exchanged between Alice and Bob are entirely hidden from Eve) such codes have to fulfill the following requirements: (i) the error probability of Bob must decrease as the code length increases (reliability constraint) and (ii) the equivocation rate of Eve must increase as the code length increases (secrecy constraint). The wiretap code can be designed based on the following rates: (i) rate of transmitted codewords, $R_B \in \mathbb{R}_+$ , and (ii) rate of transmitted confidential information (i.e., target secrecy rate), $R \in \mathbb{R}_+$. Notice that $R_B\leq C_B$ is chosen to guarantee the reliability constraint whereas $R_E>C_E$ guarantees the secrecy constraint, in which $R_E = R_B - R$ is the rate of redundancy used to confuse Eve. Observe that to achieve the maximum $R$, the complete \acp{CSI} of Bob and Eve have to be available at Alice. However, it is challenging for Alice to obtain \ac{CSI} of Eve since, in practice, Eve is passive and then $R_E>C_E$ cannot always be fulfilled (i.e., perfect secrecy is not guaranteed). In this case, Alice may choose a fixed $R$ and the secrecy outage probability turns a reasonable approach to quantify the level of security of a data communication system \cite{barros}. In this way, one may use the following achievable secrecy rate:
\begin{align}\label{eq:Rs}
	R_S &= \frac{1}{N}\left[\log_2\left[\det\left(\mathbf{I}_{N} +  \mathbf{\Lambda}_{\gamma_{B}}\right)\right]-\right. \nonumber \\
		&\hspace{1.7cm}\left.\log_2\left[\det\left(\mathbf{I}_{N} +  \mathbf{\Lambda}_{\gamma_{E}}\right)\right]\right]^+ ~[\text{bps}/\text{Hz}].
\end{align}
%As discussed in \cite{yan}, a wiretap code is necessary to obtain (\ref{eq:Cs2}) and fulfill the following requirements for having perfect secrecy: (i) the error probability of Bob must decrease as the code length increases (reliability constraint) and (ii) the equivocation rate of Eve must increase as the code length increases (secrecy constraint). The wiretap code can be designed based on the following rates: (i) rate of transmitted codewords, $R_B \in \mathbb{R}_+$ , and (ii) rate of transmitted confidential information (i.e., target secrecy rate), $R \in \mathbb{R}_+$. Notice that $R_B\leq C_B$ is chosen to guarantee the reliability constraint whereas $R_E>C_E$ guarantees the secrecy constraint, in which $R_E = R_B - R$ is the rate of redundancy used to confuse Eve. 
%Observe that to achieve the maximum $R$, the complete \acp{CSI} of Bob and Eve have to be available at Alice. However, it is challenging for Alice to obtain \ac{CSI} of Eve since, in practice, Eve is passive and then $R_E>C_E$ cannot always be fulfilled (i.e., perfect secrecy is not guaranteed). In this case, Alice may choose a fixed $R$ and the secrecy outage probability turns a reasonable approach to quantify the level of security of a data communication system \cite{barros}.
%An outage event occurs when the channel between Alice and Bob is in outage or Eve is capable of decoding private messages exchanged between Alice and Bob. 
Consequently, the secrecy outage probability can expressed as
\begin{align}\label{eq:ps}
	P_S(R) &= \mathbb{P}\left\lbrace R_S < R\right\rbrace \nonumber \\
		   &= \mathbb{P}\left\lbrace\det\left(\frac{\mathbf{I}_{N} +  \mathbf{\Lambda}_{\gamma_{B}}}{\mathbf{I}_{N} +  \mathbf{\Lambda}_{\gamma_{E}}}\right)<2^{RN}\right\rbrace.
\end{align}
Notice that perfect secrecy is achieved when $R_S>R$ whereas $R_S<R$ means that perfect secrecy is not guaranteed.
	
% --------------------------------------------------------------
% NUMERICAL RESULTS
% --------------------------------------------------------------
\begin{figure*}[!htb]
	\centering
	\psfrag{Rates (Mbps)}[c][c][.75]{$\bar{R}_S$ (Mbps)}
	\psfrag{P}[c][c][.75]{$P_T$ (dBm)}
	\psfrag{RsSP-UA}[l][l][.7]{SP~$-$~UA}
	\psfrag{RsSP-OA}[l][l][.7]{SP~$-$~OA}
	\psfrag{RsLP-UA}[l][l][.7]{LP~$-$~UA}
	\psfrag{RsLP-OA}[l][l][.7]{LP~$-$~OA}
	\psfrag{RB-UA}[l][l][.7]{$(C_E=0)~-$~UA}
	\psfrag{RB-OA}[l][l][.7]{$(C_E=0)~-$~OA}
	\psfrag{a}[c][c][.8]{(a)~$\bar{\gamma}_{B,1}\in[51.1,~61.1)$~dB.}
	\psfrag{b}[c][c][.8]{(b)~$\bar{\gamma}_{B,2}\in[61.1,~72.3)$~dB.}
	\psfrag{c}[c][c][.8]{(c)~$\bar{\gamma}_{B,3}\in[72.3,~82.9]$~dB.}
	\includegraphics[width=\linewidth]{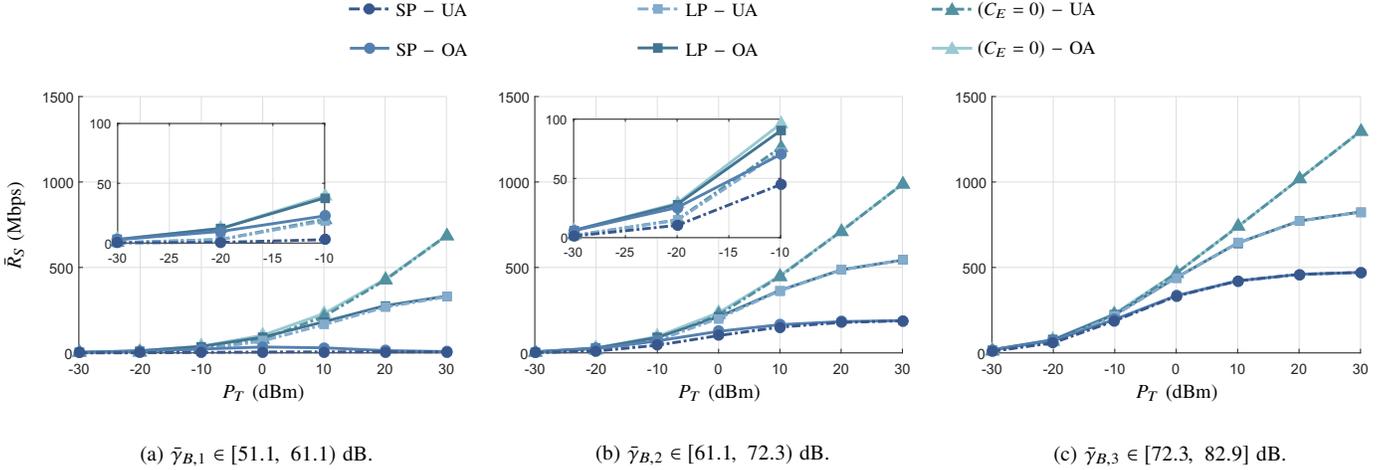}
	\caption{$\bar{R}_S\times P_T$ for the hybrid wiretap channel under the adoption of both OA and UA techniques.}
	\label{fig:Cs_Pt}
\end{figure*}
	
\section{Numerical Results}\label{Results}
This section presents numerical analysis of secrecy outage probability for the hybrid wiretap channel model assuming that Eve is passive and Alice has only \ac{CSI} of Bob. Additionally, the ergodic achievable secrecy rate $\bar{R}_S~=~B_w\mathbb{E}_{\mathbfcal{H}_{B},\mathbfcal{H}_{E}}\left[R_S\right]$ is also evaluated, where $B_w$ is the frequency bandwidth. Although theoretical, $\bar{R}_S$ is an interesting parameter to quantify the leakage information to Eve when data communication is performed between Alice and Bob. The frequency band $1.7-86$~MHz (i.e., broadband applications in the sense of the \ac{PLC} community) and $N = 2048$ apply. \acused{OA} Optimal power allocation (OA) by using the water-filling algorithm \cite{cioffi} and \ac{UA} techniques are taken into account. The total transmission power  ($P_T$) ranges from $-30$~dBm to $30$~dBm, in which the intervals $[-30,0)$~dBm and $[0,30]$~dBm cover theoretical and practical values of $P_T$, respectively.
	
It is important to state that real data is used to analyze the hybrid wiretap channel model \cite{oliveira2015,oliveira2016}. Such a data set, obtained from a measurement campaign carried out in Brazilian in-home facilities, constitutes estimates of \ac{PLC} and hybrid PLC-wireless \acp{CFR}. They are considered to represent, respectively, the Alice-Bob and Alice-Eve links. In particular, hybrid PLC-wireless \ac{CFR} estimates allow one to analyze two kinds of situation for the hybrid wiretap channel model:
\begin{itemize}
	\item \textit{Eve is close to Alice}: This situation represents the hybrid PLC-wireless \acp{CFR} measured within a $2$~meters radius centered at the outlet where the PLC transmitter (Alice) was connected.
	\item \textit{Eve is close to Bob}: This situation represents the hybrid PLC-wireless \acp{CFR} measured within a $2$~meters radius centered at the outlet where the PLC receiver (Bob) was connected.
\end{itemize}
Following \cite{oliveira2016}, the situations where Eve is close to Alice and is close to Bob are named, respectively, \textit{short-path} (SP) and \textit{long-path} (LP).
	
Table \ref{tab:nSNR} lists the maximum, mean, minimum, standard deviation (SD), $90\%$ percentile values of the \ac{nSNR} multi-channel regarding the estimates of PLC and hybrid PLC-wireless \acp{CFR}. According to \cite{cioffi}, the mathematical definition of the \ac{nSNR} multi-channel parameter is 
\begin{align}
	\bar{\gamma}_{l} \triangleq \det\left(\mathbf{I}_N + \mathbf{\Lambda}_{|\mathbfcal{H}_{l}|^2}\mathbf{\Lambda}_{P_{V_l}}^{-1}\right)^{1/N}-1.
\end{align}
From Table \ref{tab:nSNR}, observe that the maximum $\bar{\gamma}_{l}$ values of PLC, hybrid PLC-wireless SP, and hybrid PLC-wireless LP channels are, respectively, $82.8$, $69.6$, and $56.4$~dB whereas the minimum $\bar{\gamma}_{l}$ values for them are $51.1$, $54.3$, and $36.9$~dB, respectively.
\begin{table}[!htb]
	\renewcommand{\arraystretch}{1.5}
	\caption{Statistics of $\bar{\gamma}_l$ in Decibels (dB)}
	\label{tab:nSNR}
	\centering
	\arrayrulecolor[HTML]{708090}
	\setlength{\arrayrulewidth}{.1mm}
	\setlength{\tabcolsep}{4pt}
	\begin{tabular}{c|c|c|c|c|c|c}
		\hline
		\textbf{Receiver}  & \textbf{Maximum} & \textbf{Mean} & \textbf{Minimum} & \textbf{SD} & \textbf{90 \%}  \\ \hline
		PLC ($\bar{\gamma}_{B}$)  &    $82.8$       &    $70.2$    &      $51.1$     &   $9.3$    &    $81.2$      \\ \hline
		Hybrid SP ($\bar{\gamma}_{E}$)       &    $69.6$       &    $61.1$    &      $54.3$     &   $2.5$    &    $64.1$      \\ \hline
		Hybrid LP ($\bar{\gamma}_{E}$)       &    $56.4$       &    $47.2$    &      $36.9$     &   $3.8$    &    $51.9$      \\ \hline
	\end{tabular}
\end{table}
Notice that as Bob moves away from Alice, $\bar{\gamma}_{B}$ decreases and, as a consequence, it may jeopardize \ac{PLS}. On the other hand, if Bob is close to Alice, higher values of $\bar{\gamma}_{B}$ are observed, making the decoding of the information exchanged between Alice and Bob a hard task to be accomplished by Eve. In this regard and for the sake of simplicity, the assessment of numerical results considers three distinct ranges of $\bar{\gamma}_{B}$: $\bar{\gamma}_{B,1}\in[51.1,~61.1)$~dB, $\bar{\gamma}_{B,2}\in[61.1,~72.3)$~dB, and $\bar{\gamma}_{B,3}\in[72.3,~82.9]$~dB.

\begin{figure*}[!htb]
	\centering
	\psfrag{Pout}[c][c][.75]{$P_S(R)$}
	\psfrag{R}[c][c][.75]{Target secrecy rate $R$ (bps/Hz)}
	\psfrag{SP0}[l][l][.7]{SP, $P_T = -30$~dBm}
	\psfrag{LP0}[l][l][.7]{LP, $P_T = -30$~dBm}
	\psfrag{SP10}[l][l][.7]{SP, $P_T = 0$~dBm}
	\psfrag{LP10}[l][l][.7]{LP, $P_T = 0$~dBm}
	\psfrag{SP20}[l][l][.7]{SP, $P_T = 30$~dBm}
	\psfrag{LP20}[l][l][.7]{LP, $P_T = 30$~dBm}
	\psfrag{a}[c][c][.8]{(a)~OA and $\bar{\gamma}_{B,1}\in[51.1,~61.1)$~dB.}
	\psfrag{b}[c][c][.8]{(b)~OA and $\bar{\gamma}_{B,2}\in[61.1,~72.3)$~dB.}
	\psfrag{c}[c][c][.8]{(c)~OA and $\bar{\gamma}_{B,3}\in[72.3,~82.9]$~dB.}
	\psfrag{d}[c][c][.8]{(d)~UA and $\bar{\gamma}_{B,1}\in[51.1,~61.1)$~dB.}
	\psfrag{e}[c][c][.8]{(e)~UA and $\bar{\gamma}_{B,2}\in[61.1,~72.3)$~dB.}
	\psfrag{f}[c][c][.8]{(f)~UA and $\bar{\gamma}_{B,3}\in[72.3,~82.9]$~dB.}
	\includegraphics[width=\linewidth]{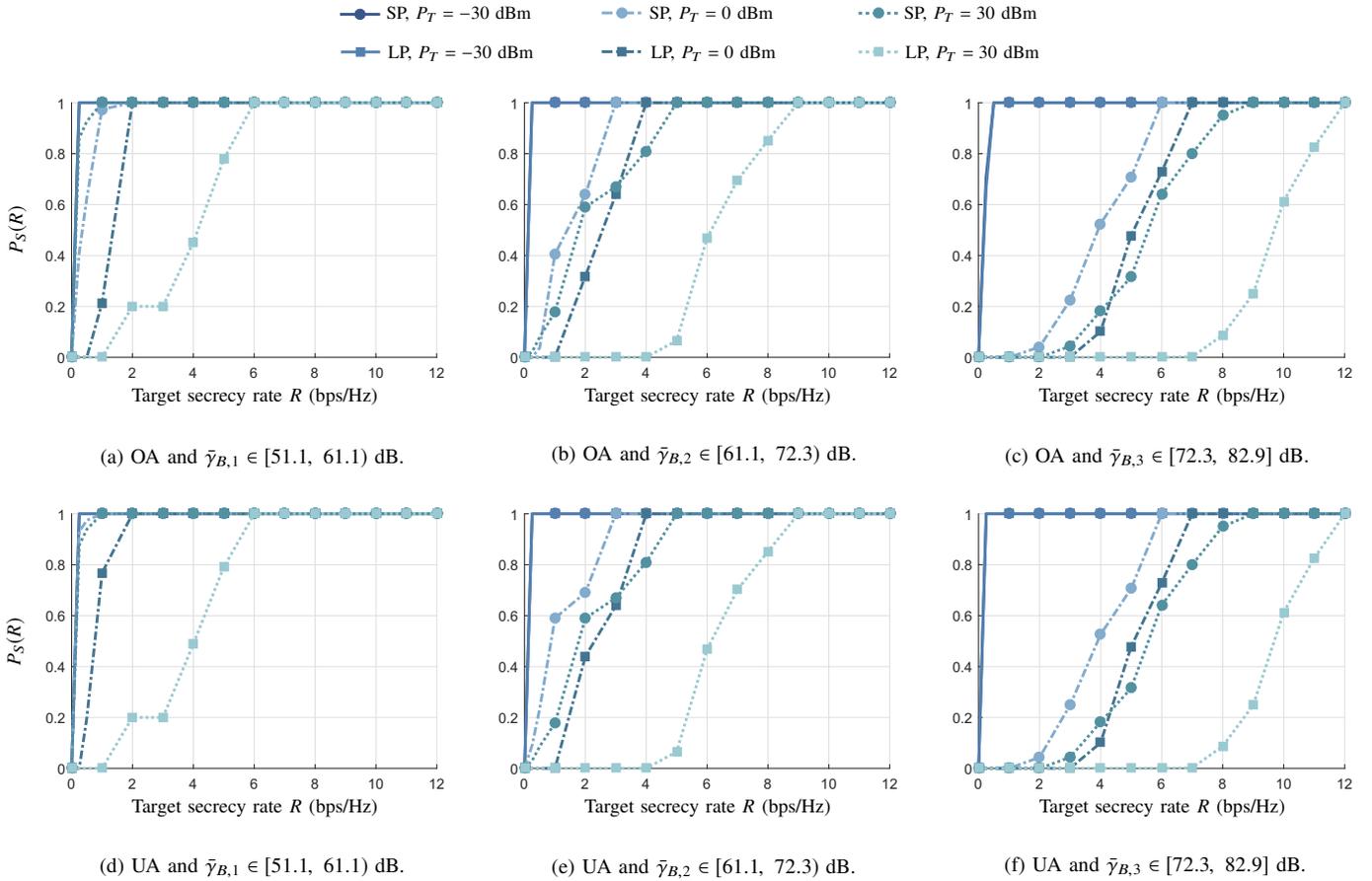}
	\caption{$P_S(R)\times R$ with $P_T\in\{-30,0,30\}$~dBm for the hybrid wiretap channel.}
	\label{fig:Pout}
\end{figure*}
	
Figs. \ref{fig:Cs_Pt}(a), (b), and (c) show $\bar{R}_S \times P_T$ for the hybrid wiretap channel model under the adoption of \ac{OA} and \ac{UA} together with, respectively, $\bar{\gamma}_{B,1}$, $\bar{\gamma}_{B,2}$, and $\bar{\gamma}_{B,3}$. In addition to that, the scenario having Eve far away from Alice and Bob (i.e., $C_E=0$) account for a reference to analyze the leakage information from the channel between Alice and Bob to Eve. Figs. \ref{fig:Cs_Pt}(a), (b), and (c) show that the difference between \ac{OA} and \ac{UA} is minimal regardless of $P_T$ and the distance between Alice and Bob. Also, notice that as Bob moves away from Alice, $R_S$ decreases in both SP and LP scenarios, i.e., more information is leaked to Eve. In particular, for $\bar{\gamma}_{B,1}$ (see Fig. \ref{fig:Cs_Pt}(a)), the SP scenario has $\bar{R}_S$ around zero for \ac{UA}, regardless of $P_T$. For instance, $\bar{R}_S$ achieves only $30$~kbps when $P_T=-10$~dBm and \ac{OA} are adopted. Also, the differences between SP and LP scenarios, in terms of $\bar{R}_S$, are almost the same in Figs. \ref{fig:Cs_Pt}(a), (b), and (c), achieving the values $322.9$, $355.2$, and $353.7$~kbps, respectively, for $P_T=30$~dBm. Lastly, considering $P_T=30$~dBm, the $\bar{R}_S$ values of SP, LP, and $C_E=0$ scenarios are equal to, respectively, $5.84$, $328.7$, and $685.8$~kbps in Fig. \ref{fig:Cs_Pt}(a); $187.6$, $542.8$, and $987.9$~kbps in Fig. \ref{fig:Cs_Pt}(b); and $471.0$, $824.7$, and $1,296$~kbps in Fig. \ref{fig:Cs_Pt}(c). For the practical values of total transmission power (i.e., $0\leq P_T\leq30$~dBm), $\bar{R}_S>0$ occurs for $\bar{\gamma}_{B,2}$ and $\bar{\gamma}_{B,3}$ in both SP and LP scenarios. In contrast, if $\bar{\gamma}_{B,1}$ applies, values of $\bar{R}_S$ around zero arise, mainly with \ac{UA}.

Fig. \ref{fig:Pout} depicts $P_S\times R$ for the hybrid wiretap channel model considering $P_T\in\{-30,0,30\}$~dBm. In this regard, OA (see Figs. \ref{fig:Pout}(a), (b), and (c)) and UA (see Figs. \ref{fig:Pout}(d), (e), and (f)) are presented. Also, $\bar{\gamma}_{B,1}$, $\bar{\gamma}_{B,2}$, and $\bar{\gamma}_{B,3}$ are considered in Figs. \ref{fig:Pout}(a), (b), and (c), respectively, and in Figs. \ref{fig:Pout}(d), (e), and (f), respectively. The figures show a minimal difference between \ac{OA} and \ac{UA} so that a small advantage for \ac{OA} can be observed in $\bar{\gamma}_{B,1}$ and $\bar{\gamma}_{B,2}$ cases when $P_T$ is equal to $-30$ and $0$~dBm. Still in Fig. \ref{fig:Pout}, notice that as Bob moves away from Alice, $P_S(R)$ increases significantly. In particular, considering \ac{OA}, $P_T=30$~dBm, and $R=6$~bps/Hz, SP and LP scenarios present $P_S(R)$ equal to $1$ for $\bar{\gamma}_{B,1}$. Also, SP and LP scenarios show $P_S(R)$ around $0.64$ and equal to $0$, respectively, for $\bar{\gamma}_{B,3}$ and $P_S(R)$ equal to $1$ and around $0.47$, respectively, for $\bar{\gamma}_{B,2}$. Finally, observe that when $P_T=-30$~dBm and $R\geq1$~bps/Hz, $P_S(R)$ is equal to $1$ in both SP and LP scenarios regardless of the distance between Alice and Bob and the used power allocation technique. Finally, as shown in Figs. \ref{fig:Pout}(a) and (d), the practical values of total transmission power (i.e., $0$ and $30$~dBm) result in high values of $P_S(R)$ for the SP scenario. 
	
\begin{figure*}[!htb]
	\centering
	\psfrag{Pout}[c][c][.75]{$P_S(R)$}
	\psfrag{P}[c][c][.75]{$P_T$ (dBm)}
	\psfrag{SP0}[l][l][.7]{SP, $R = 0.25$~bps/Hz}
	\psfrag{LP0}[l][l][.7]{LP, $R = 0.25$~bps/Hz}
	\psfrag{SP10}[l][l][.7]{SP, $R = 0.50$~bps/Hz}
	\psfrag{LP10}[l][l][.7]{LP, $R = 0.50$~bps/Hz}
	\psfrag{SP20}[l][l][.7]{SP, $R = 1.00$~bps/Hz}
	\psfrag{LP20}[l][l][.7]{LP, $R = 1.00$~bps/Hz}
	\psfrag{a}[c][c][.8]{(a)~OA and $\bar{\gamma}_{B,1}\in[51.1,~61.1)$~dB.}
	\psfrag{b}[c][c][.8]{(b)~OA and $\bar{\gamma}_{B,2}\in[61.1,~72.3)$~dB.}
	\psfrag{c}[c][c][.8]{(c)~OA and $\bar{\gamma}_{B,3}\in[72.3,~82.9]$~dB.}
	\psfrag{d}[c][c][.8]{(d)~UA and $\bar{\gamma}_{B,1}\in[51.1,~61.1)$~dB.}
	\psfrag{e}[c][c][.8]{(e)~UA and $\bar{\gamma}_{B,2}\in[61.1,~72.3)$~dB.}
	\psfrag{f}[c][c][.8]{(f)~UA and $\bar{\gamma}_{B,3}\in[72.3,~82.9]$~dB.}
	\includegraphics[width=\linewidth]{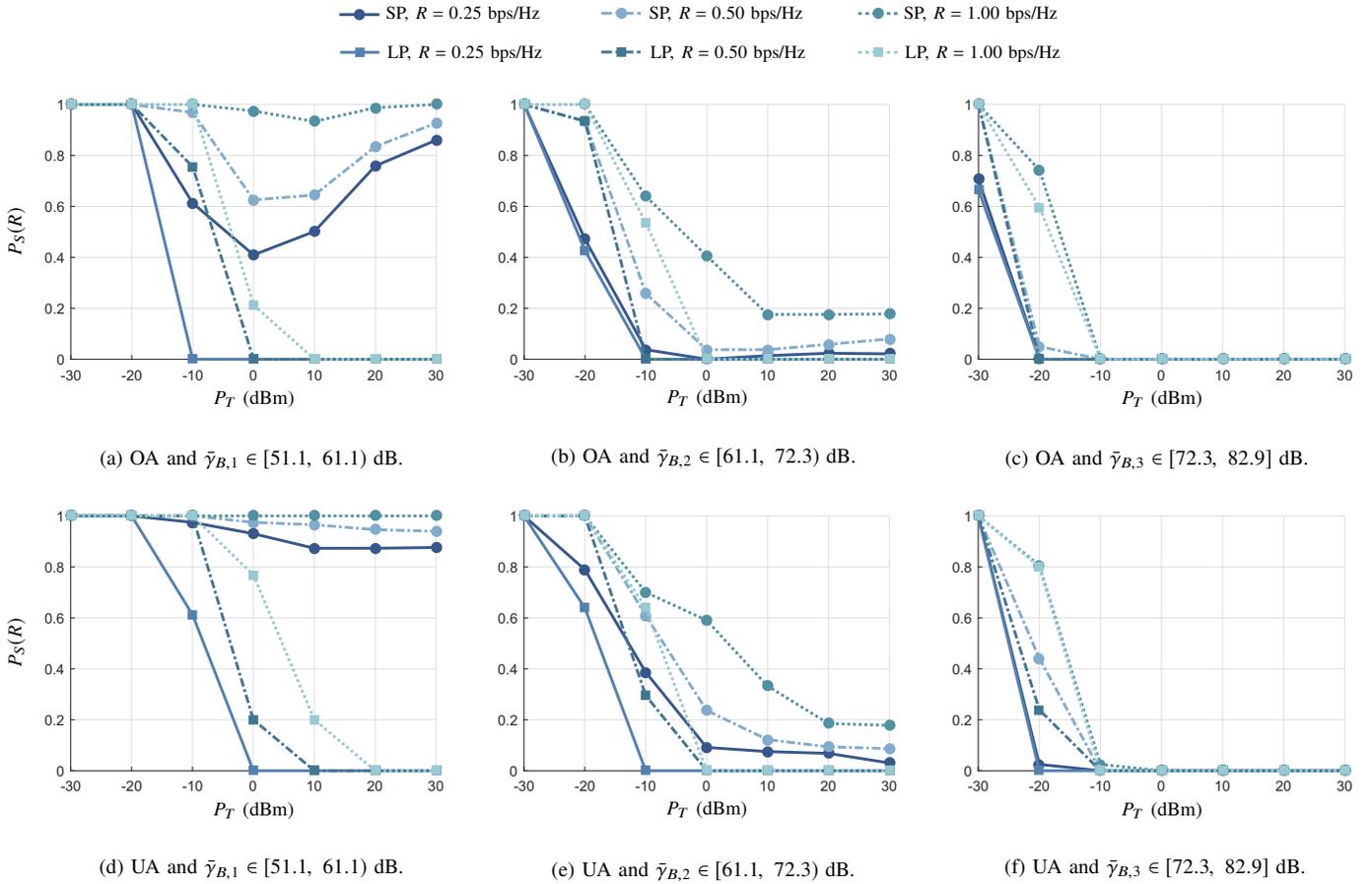}
	\caption{$P_S(R)\times P_T$ with $R\in\{0.25,0.50,1.00\}$~bps/Hz for the hybrid wiretap channel.}
	\label{fig:PoutR}
\end{figure*}

Fig. \ref{fig:PoutR} shows $P_S(R) \times P_T$ for the hybrid wiretap channel model considering $R\in\{0.25,0.50,1.00\}$~bps/Hz. Also, \ac{OA} (see Figs. \ref{fig:PoutR}(a), (b), and (c)) and \ac{UA} (Figs. \ref{fig:PoutR}(d), (e), and (f)) are taken into account. In addition, $\bar{\gamma}_{B,1}$, $\bar{\gamma}_{B,2}$, and $\bar{\gamma}_{B,3}$ are considered in Figs. \ref{fig:PoutR}(a), (b), and (c), respectively, and in Figs. \ref{fig:PoutR}(d), (e), and (f), respectively. Notice that, as $P_T$ increases the difference between \ac{OA} and \ac{UA} tends to zero. On the other hand, if $P_T$ decreases, then a small improvement in favor of \ac{OA} occurs. Considering the LP scenario, \ac{OA}, and practical values of total transmission power, i.e., $0\leq P_T\leq30$~dBm, $P_S(R)$ is close to zero for $\bar{\gamma}_{B,1}$, $\bar{\gamma}_{B,2}$, and $\bar{\gamma}_{B,3}$ regardless of $R$, except when $P_T=0$~dBm in the $\bar{\gamma}_{B,1}$ case, in which $P_S(R)$ is around $0.2$.
Concerning the SP scenario, if \ac{OA} and $0\leq P_T\leq30$~dBm apply, then $P_S(R)$ is close to zero for $\bar{\gamma}_{B,3}$ regardless of $R$ and for $\bar{\gamma}_{B,2}$ when $R$ is equal to $0.25$~bps/Hz. Besides, $R=0.50$ and $1.00$~bps/Hz provide $P_S(R)<1$  and $P_S(R)\leq0.4$, respectively, for $\bar{\gamma}_{B,2}$. Lastly, observe that $P_S(R)$ values higher than $0.40$, $0.60$ and $0.90$ are found when $R$ is equal to $0.25$, $0.50$, and $1.00$ bps/Hz, respectively, for $\bar{\gamma}_{B,1}$.
	
Overall, the attained results show that Eve may be able to eavesdrop confidential information exchanged between Alice and Bob in the SP scenario together with $\bar{\gamma}_{B,1}$, which account for the situations where Alice is far from Bob (i.e., the distance is around $6$ meters) and Eve is close to Alice (i.e., the distance is shorter than $2$ meters), respectively. Moreover, regardless of $R$, the use of practical or theoretical values of $P_T$ results in high values of $P_S(R)$, mainly when \ac{UA} is adopted. Such finding shows that the radiation of the \ac{PLC} signal could jeopardize \ac{PLS} if Eve is a wireless device operating in the same frequency band and located close to Alice, which constitutes a typical scenario associated with electric power grids composed of unshielded power cables in in-home facilities. 

% --------------------------------------------------------------
% CONCLUSION
% --------------------------------------------------------------
\section{Conclusion}\label{Conclusion}
This study has quantitatively investigated in a real scenario how secure a broadband in-home \ac{PLC} system can be when a passive wireless device, operating in the vicinity of power cables, overhears private messages exchanged between two \ac{PLC} devices. In this regard, with the adoption of the hybrid wiretap channel model, mathematical formulations of achievable secrecy rate and secrecy outage probability have been provided and numerically analyzed. 

The numerical results have shown how vulnerable a broadband in-home \ac{PLC} system can be when unshielded power cables constitute electric power grids and a malicious wireless device is located in the vicinity of them. In the worst scenario, where Bob is far from Alice while Eve is close to Alice (i.e., the distance is shorter than $2$ meters), high values of secrecy outage probability arise for all analyzed values of target secrecy rate and total power transmission (practical and theoretical values).

Overall, the discussed results reinforce the importance and necessity of the introduction of novelties in the design of \ac{PLC} devices or the use of shielded power cables in the electrical power grid for performing both energy delivery and data communication when the discussion related to the physical layer of \ac{PLC} systems involves security issues.

%\section*{Acknowledgments}

\ifCLASSOPTIONcaptionsoff
\newpage
\fi


\end{document}